\documentclass[pre,floatfix,twocolumn]{revtex4}

\usepackage{amsmath,amssymb,latexsym,epsfig,graphics,epsf}
\usepackage{graphics}
\usepackage{float}

\usepackage{verbatim}
\usepackage{times}
\usepackage{amsmath}

\newcommand{\Sex}{S_{\textrm{ex}}}

\newcommand{\bfa}[1]{\mathbf{#1}}

\begin{document}

\title{Isomorph invariance of dynamics of sheared glassy systems}
\date{\today}
\author{Yonglun Jiang}
\author{Eric R. Weeks}
\affiliation{Physics Dept., Emory University, 400 Dowman Dr, Atlanta, GA 30322, USA.}
\author{Nicholas P. Bailey}
\affiliation{DNRF Center ``Glass and Time'', IMFUFA, Dept. of Science and Environment, Roskilde
 University, P. O. Box 260, DK-4000 Roskilde, Denmark}

\begin{abstract}
  We study hidden scale invariance in the glassy phase of the Kob-Andersen binary Lennard-Jones system. After cooling below the glass transition, we generate a so-called isomorph from the fluctuations of potential energy and virial in the NVT ensemble -- a set of density, temperature pairs for which structure and dynamics are identical when expressed in appropriate reduced units. To access dynamical features we shear the system using the SLLOD algorithm coupled with Lees-Edwards boundary conditions, and study the statistics of stress fluctuations and the  particle displacements transverse to the shearing direction. We find good collapse of the statistical data showing that isomorph theory works well in this regime. The analysis of stress  fluctuations, in particular the distribution of stress changes over a given strain interval, allows us to identify a clear signature of avalanche behavior in the form of an exponential tail on the negative side. This feature is also isomorph invariant. The implications of isomorphs for theories of plasticity are discussed briefly.
\end{abstract}

\maketitle

\section{Introduction}

Recently it has been discovered that a broad class of classical condensed matter systems exhibit an approximate scale invariance\cite{Bailey/others:2008a,Bailey/others:2008b,Bailey/others:2008c,Schroder/others:2009b,Gnan/others:2009,Schroder/others:2011}. Upon changing a system's density, a corresponding change in temperature can be found such that the structure and dynamics of the system are unchanged -- as long as they are compared in an appropriate dimensionless form. State points which are equivalent in this sense are said to be {\em isomorphic}, and the key feature of systems exhibiting so-called hidden scale invariance is the existence of isomorphic curves, or isomorphs, in the phase diagram\cite{Gnan/others:2009}. The theory of isomorphs shows how they can be identified straightforwardly in computer simulations, how to appropriately scale quantities for comparison, and which quantities are expected to be isomorph-invariant. Isomorphs have been identified and investigated in the equilibrium liquid state for many model systems\cite{Bailey/others:2008a,Ingebrigtsen/Schroder/Dyre:2012a,Ingebrigtsen/Schroder/Dyre:2012b,Boehling/others:2012,Veldhorst/Dyre/Schroder:2014}; they have been studied in conditions of non-equilibrium steady-state shearing\cite{Separdar/others:2013} and aging\cite{Gnan/others:2009,Gnan/others:2010, Dyre:2018} and zero temperature shearing of a glass\cite{Lerner/Bailey/Dyre:2014}.  The class of systems exhibiting good isomorphs has been denoted ``R-simple systems''. For reviews the reader may consult Refs.~\onlinecite{Ingebrigtsen/Schroder/Dyre:2012a,Dyre:2014,Dyre:2016}. Isomorphs have not, however, been investigated in the context of deformation of the glass state at finite temperature; that is the topic of this work.

We consider an amorphous solid created by cooling a viscous liquid down below its glass transition and then applying Couette-type shearing at constant volume and fixed strain rate. This necessarily involves a departure from equilibrium and in principle introduces a potential dependence on history, for example through cooling rate, as well as possible aging effects, into the system's behavior. We minimize these issues by restricting our attention to steady state shearing: if one shears the system at a constant strain rate beyond say 50\% or 100\% strain, a steady state is obtained which depends only on the density, the temperature, and the strain rate. As discussed in Ref.~\onlinecite{Separdar/others:2013}, the existence of isomorphs reduces these three variables to two: a variable labeling the isomorph (in equilibrium this is generally taken to be the excess entropy) and a dimensionless strain rate. In principle, however, isomorph theory allows for independent configurations from equilibrium states above the glass transition which are isomorphic to each to be cooled into the glassy state in an isomorphic way, such that the entire thermal histories and deformation histories are isomorphic. In that case the entire stress-strain curves could be compared, rather than simply the steady state part. Some ten years ago Lerner and Procaccia proposed a scaling theory for steady state plasticity based on approximating the pair potential by an inverse power law\cite{Lerner/Procaccia:2009a}. The relation between that work and isomorph theory will be discussed below.

We work with the usual Kob-Andersen binary Lennard-Jones glass forming model\cite{Kob/Andersen:1994, Kob/Andersen:1995a, Kob/Andersen:1995b}, which is useful because it is difficult (though not impossible\cite{Toxvaerd/others:2009,Ingebrigtsen/others:2018}) to crystallize on computer time scales. It is certainly straightforward to obtain a glassy state in a simulation with sufficiently rapid cooling. We consider two starting states in the glass, one just below the glass transition and one deep in the glass. The focus of the analysis is on analyzing steady-state stress strain curves statistically and the particle displacements characterized by the mean squared displacement (MSD).

Demonstrating the presence of good isomorphs in the glassy state has theoretical relevance not just because it permits a simplification of the phase diagram, but for two other reasons. First, given the existence of isomorphs, it becomes clearer what the relevant thermodynamic variables are: pressure, while being of course extremely relevant from an experimental point of view, becomes secondary to density. Moreover strain rates should  be specified and compared in their dimensionless (reduced) form. Secondly, the existence of isomorphs puts a strong constraint on theories of glassy behavior. Several theories for the mechanical properties of amorphous materials have been proposed. Hidden scale invariance imposes constraints on candidate theories, since a theory which purports to be general should in particular apply to systems with hidden scale invariance, and should therefore involve equations expressed in reduced-unit quantities which are explicitly isomorph invariant. This principle was called the ``isomorph filter'' in the context of theories of the glass transition\cite{Gnan/others:2009}.

The following section gives an overview of the most essential results from the theory of isomorphs. Section~\ref{sec:simulations} then describes the system and the simulation methods used. Section~\ref{sec:glass_isomorphs} describes how we generated glassy isomorphs and checks the isomorph invariance of their structure using the radial distribution function. The main analysis of the paper is presented in the following two sections; Sec.~\ref{sec:stress_strain} presents a detailed analysis of the stress strain curves while Sec.~\ref{sec:particle_dymamics} contains an analysis of particle displacements via the mean squared displacement transverse to the shearing direction. Section~\ref{discussion} discusses implications of the existence of isomorphs for theories of plasticity, showing via an example from the literature how density dependence can be included  in an isomorph invariant way. Section~\ref{conclusion} summarizes and concludes the paper.

\section{Isomorph theory}

In this section we give a brief overview of the theoretical basis for analyzing isomorphs, starting with how to put observables in the necessary dimensionless forms needed to properly compare structure and dynamics at different thermodynamic state points.

\subsection{\label{reduced_units}Reduced units}

As mentioned above, quantities must be expressed in an appropriate dimensionless form, referred to as ``using reduced units.'' We essentially scale out the direct effects of changing density and temperature on structure and dynamics: If we have $N$ particles in a volume $V$ then the system's (number) density is $ \rho\equiv N/V$. A basic length scale $l_0$ is defined by by interparticle spacing $\rho^{-1/3}$. If the system is in equilibrium at temperature $T$ then a basic time scale is defined by the time for a particle with the thermal velocity $\sqrt{k_BT/m}$ to cover a distance  $l_0$: $t_0=\rho^{-1/3} (k_BT/m)^{-1/2}$. In the case of a mixture, the average mass $\langle m\rangle$ should be used. Given $l_0$, $t_0$ we can rescale space and time, making it possible, for example, to compare trajectories at different state points --- the rescaling accounts for the most trivial effects of changing density and temperature. In fact all physical quantities can be rescaled similarly, by taking appropriate combinations of $l_0$, $t_0$ and $\langle m  \rangle$. For a quantity with dimensions of energy the scale factor is just $k_BT$. For a pressure (or stress, or elastic modulus) the scale factor is $\rho k_BT$. We denote the rescaled, ``reduced-unit'', quantities with a tilde, thus the reduced form of a particle position $\bfa{r}$ is  $\bfa{\tilde r}\equiv \rho^{1/3} \bfa{r}$.

\subsection{Identifying isomorphs}

The scale invariance that underlies the existence of isomorphs derives ultimately from the fact that the potential energy surface of the $N$-particle system changes in a somehow homogeneous way when density is changed. For example suppose changing density of any microscopic configuration by a factor $\lambda$ results in the potential energies being changed by a factor $\lambda^\gamma$ for some exponent $\gamma$. This can then be compensated by increasing temperature by the same factor, meaning all Boltzmann factors will be unchanged, so the statistical probability of all microstates will be the same at the new density as for the corresponding unscaled configurations at the original density. It follows that all statistical measures of structure will be invariant when expressed in terms of the reduced coordinates  $\bfa{\tilde r}$. It can also be shown\cite{Schroder/others:2009b} that the equation of motion is also the same for both states when expressed in reduced units, and that therefore all dynamical quantities are also invariant in reduced units. The case just described is realized by systems interacting with an inverse power law (IPL) pair potential $v=A/r^n$; in that case the scaling exponent $\gamma$ is given by $n/3$. In that case isomorphs are exact, trivial, well-known and the invariant quantities include not just all structural and dynamical but also thermodynamic quantities in reduced units. More generally we do not expect to find exact isomorphs, but we find very good approximations. The simplest way to express and identify hidden scale invariance was shown in Ref.~\onlinecite{Schroder/Dyre:2014}, where the essential condition was stated as follows: a change of density must preserve the order of potential energies of microstates. To test for scale invariance we consider infinitesimal changes of density, whereupon changes in the potential energies $U$ of microstates are given by

\begin{equation}\label{dU_W_d_ln_rho}
d U = W d\ln\rho;
\end{equation}
here $W$ is the virial, a quantity typically calculated in computer simulations due to its appearance in the formula for pressure\cite{Allen/Tildesley:1987}. Requiring that the order of energies be preserved means in particular that configurations at a given density with the same $U$ will experience the same change in $U$ upon an infinitesimal change of $\rho$. By Eq.~\eqref{dU_W_d_ln_rho} this means they have the same $W$. In other words, potential energy and virial must be strongly correlated (the discovery of strong $U,W$ correlations\cite{Pedersen/others:2008c} marked the beginning of the development of isomorph theory). Linear regression applied to a scatter plot of $W$ versus $U$ yields two parameters, namely the correlation coefficient

\begin{equation}
  R = \frac{\langle \Delta U \Delta W \rangle}
  {\sqrt{\langle (\Delta U)^2\rangle}  \sqrt{\langle (\Delta W)^2\rangle}},
\end{equation}
and the slope
\begin{equation}\label{gamma_fluctuations}
  \gamma = \frac{\langle \Delta U \Delta W \rangle}
  {\langle (\Delta U)^2\rangle}.
\end{equation}
Here angle brackets denote canonical ensemble averages. A value of $R$ close to unity in a region of the phase diagram (typically values above around 0.9 are considered good, although lower thresholds have also been used\cite{Hummel/others:2015}) indicates that the system exhibits hidden scale invariance and should have good isomorphs in that part of the phase diagram. The interpretation of the slope $\gamma$ was given in Ref.~\onlinecite{Gnan/others:2009}: it is the slope of curves of constant excess entropy (that is, configurational adiabats) in the $\ln\rho, \ln T$ phase diagram:

\begin{equation}\label{conf_adiabat_gamma}
  \left( \frac{d\ln T}{d\ln\rho}\right)_{\Sex} = \gamma(\rho, T).
\end{equation}
The excess entropy is defined as the entropy minus that of the ideal gas with the same density and temperature, and is one of the thermodynamic properties which is invariant along an isomorph. In the 2014 formulation of the theory the status of configurational adiabats was raised such that these are considered to define isomorphs\cite{Schroder/Dyre:2014}. Since $\gamma(\rho, T)$ can be calculated at any state point using the fluctuation formula Eq.~\eqref{gamma_fluctuations}, Eq.~\eqref{conf_adiabat_gamma} provides a general method to generate isomorphs by numerical integration. Typically steps of order 1\% in density are used.

\section{\label{sec:simulations}Simulations}

The system studied is the usual binary Lennard-Jones system introduced by Kob and Andersen\cite{Kob/Andersen:1994, Kob/Andersen:1995a, Kob/Andersen:1995b}, which has been mainly studied at one particular density, 1.2 $\sigma_{AA}^{-3}$ (where the A particles are the larger ones). From now on, when not using reduced units, we work with the unit system defined by the Lennard-Jones parameters of the A particles' interactions with each other, $\sigma_{AA}$ and $\epsilon_{AA}$, and the mass which is the same for both A and B particles; thus, temperature is given in units of $\epsilon_{AA}/k_B$. The potential is cut off using the shifted-force method\cite{Toxvaerd/Dyre:2011} at 2.5 $\sigma$ for each type of interaction. The number of particles is 1000 (with the usual composition of 80\% A). The simulations are carried on a GPU cluster using the RUMD software\cite{Bailey/others:2017, rumd:2018}.

\begin{figure}
  \includegraphics[width=0.5\textwidth]{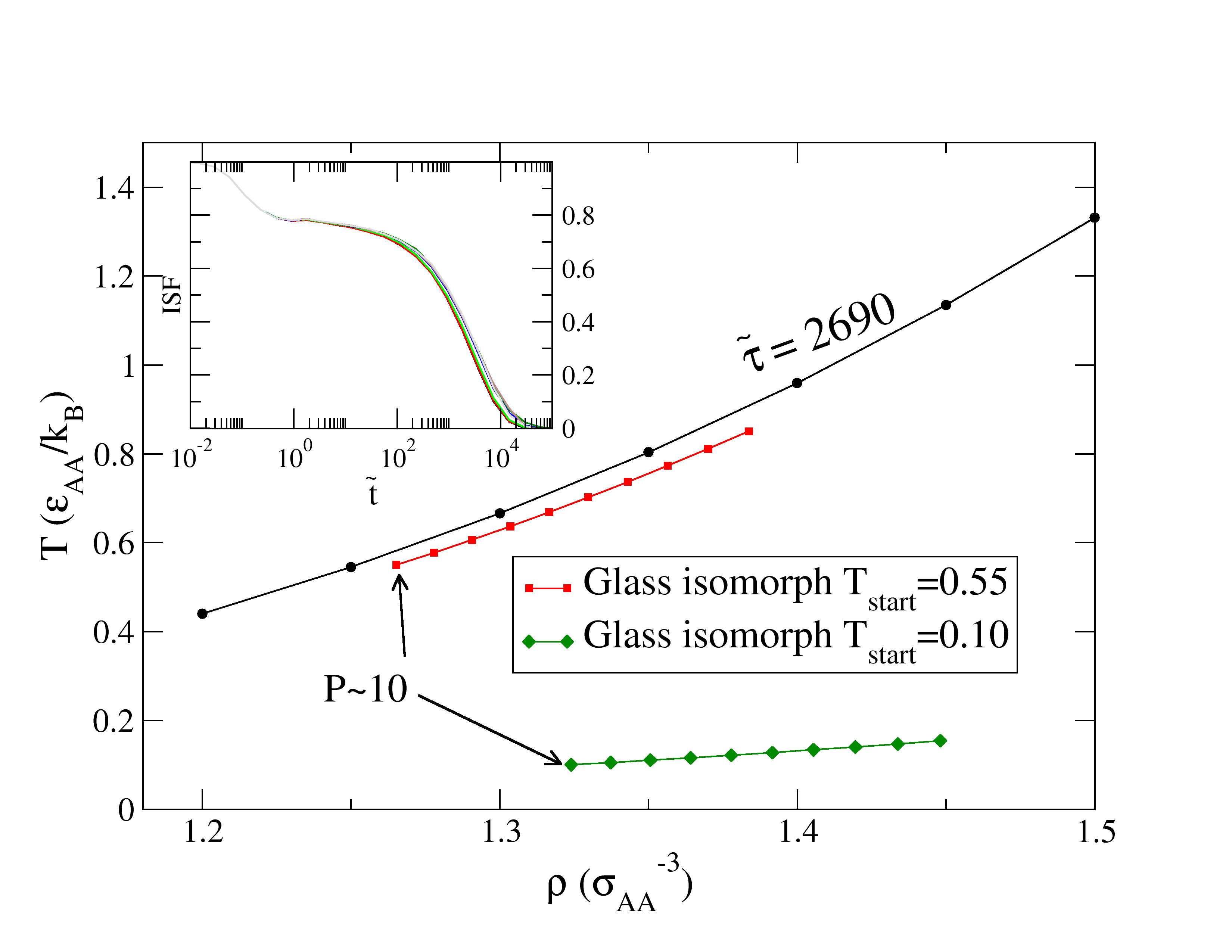}  
\caption{\label{glass_isomorph} The black symbols indicate an isomorph in the supercooled liquid which includes the point $\rho=1.2, T=0.44$. We use this as a guide to locating the glass transition; its relaxation time is about 2690 in reduced units, corresponding to 3820 in LJ units at the lowest density 1.2). The inset shows the intermediate scattering function for the different state points, all lying on top of each other. The red and green symbols indicate isomorphs generated in the glass which we use for studying deformation, referred to as those starting at temperature T=0.55 and T=0.1, respectively. Note that the starting densities are not the same, since these are taken from a cooling run at fixed pressure $P=10$.}
\end{figure}

The glassy states are created by cooling a liquid at constant pressure at a fixed cooling rate from temperature $T=1.0$ down to a given start temperature. Different cooling rates are applied, but for the steady-state results presented in this work the cooling rate is not relevant. The reason for cooling at fixed pressure rather than fixed volume is to avoid arriving at a state where the pressure is very low or negative, since good isomorphs are generally obtained at not too low pressures. To locate our glassy isomorphs in the phase diagram and compare to other work on this system, it is useful to have an idea of where the glass transition is. When considering the full phase diagram the glass transition  can be defined as the set of $\rho,T$ points where the liquid's relaxation time attains some fixed value. There are two sources of ambiguity or arbitrariness in such a definition: which observable to use when defining the relaxation time, and which value to set as defining $T_g(\rho)$. Experimentally for the latter one chooses conventionally a value of order 100 s in real units -- with the isomorph theory in mind it is natural to specify a criterion in reduced units, since in a system with isomorphs the glass line will then correspond to an isomorph\cite{Gnan/others:2010}. In computer simulations relaxation times of order 100 s are nowhere near realistic so as a guide we choose a viscous liquid state which can be equilibrated in reasonable time. In Fig.~\ref{glass_isomorph} we plot a viscous liquid isomorph whose temperature at the usual Kob-Andersen density 1.2 is 0.44. The relaxation time there (based on fitting the self-intermediate scattering function of the A particles to a stretched exponential function) is 3820 (LJ units), which corresponds to about 2690 in reduced units. This isomorph is generated using the analytical expression for Lennard-Jones potentials as described in Ref.~\onlinecite{Boehling/others:2012} (using the same reference density 1.6 but a slightly lower value of $\gamma$ at the reference density, 4.57) and simulated for $10^8$ time steps per state point.

\begin{table}
  \caption{\label{tab:isomorph_data}Thermodynamic data for the starting points of glassy isomorphs, obtained by cooling at constant pressure $P=10.0$ from temperature 1.0 over 10$^8$ steps of size $dt=0.0025$. The cooling rate is therefore 1.8 $\times$ 10$^{-6}$ for cooling to $T$=0.55 and 3.6$\times$ 10$^{-6}$ for cooling to $T$=0.1.}
  \begin{tabular} {ccccc|ccccc}\toprule
    \multicolumn{5}{l}{$T_{start}=0.55$} & \multicolumn{5}{l}{$T_{start}=0.10$}\\
    $\rho$ & $T$ & $P$ & $R$ & $\gamma$ & $\rho$ & $T$ & $P$ & $R$ & $\gamma$\\ \colrule
    1.265 & 0.550 & 9.35 & 0.955 & 4.950 & 1.324 & 0.100 & 9.75 & 0.824 & 5.011 \\
    1.278 & 0.577 & 10.68 & 0.954 & 4.971 &  1.337 & 0.105 & 11.21 & 0.834 & 5.002 \\
    1.291 & 0.606 & 11.99 & 0.962 & 5.078 & 1.351 & 0.110 & 12.79 & 0.843 & 4.953 \\
    1.304 & 0.637 & 13.72 & 0.960 & 5.033 & 1.364 & 0.116 & 14.48 & 0.855 & 4.944 \\
    1.317 & 0.669 & 15.37 & 0.965 & 5.015 & 1.378 & 0.121 & 16.29 & 0.864 & 4.916 \\
    1.330 & 0.702 & 16.99 & 0.968 & 4.936 & 1.392 & 0.127 & 18.22 & 0.873 & 4.879 \\
    1.343 & 0.737 & 18.94 & 0.972 & 4.927 & 1.406 & 0.134 & 20.29 & 0.879 & 4.873 \\
    1.356 & 0.773 & 21.07 & 0.973 & 4.874 & 1.420 & 0.140 & 22.49 & 0.886 & 4.829 \\
    1.370 & 0.811 & 23.09 & 0.976 & 4.901 & 1.434 & 0.147 & 24.85 & 0.893 & 4.817 \\
    1.384 & 0.851 & 25.24 & 0.979 & 4.869 & 1.448 & 0.154 & 27.37 & 0.899 & 4.799 \\ \botrule
  \end{tabular}
  
\end{table}

\section{\label{sec:glass_isomorphs}Glass isomorphs}
  
\begin{figure}
  \includegraphics[width=0.5\textwidth]{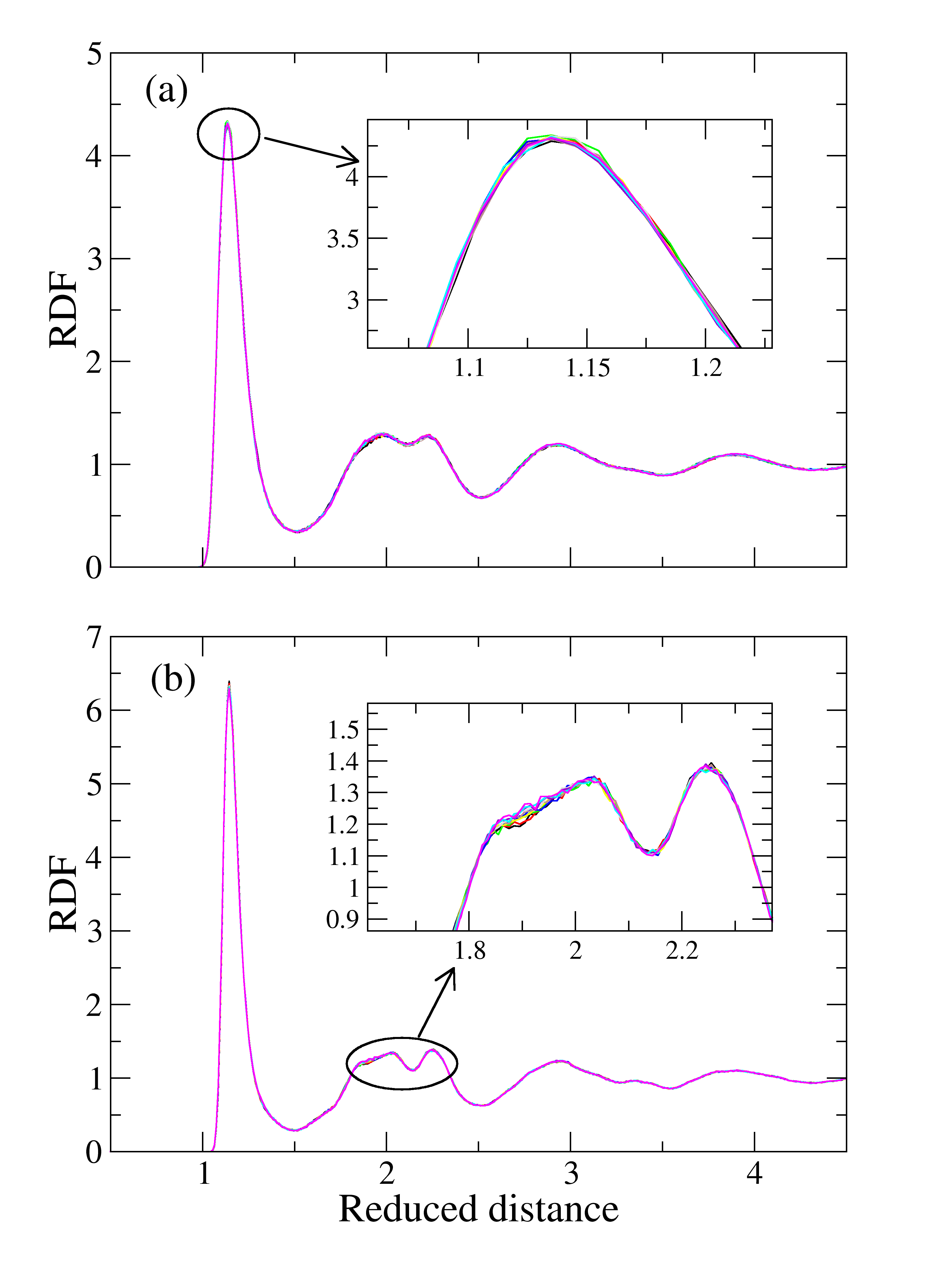}
  \caption{\label{glassy_rdf} Radial distribution function for the large (A) particles in reduced units along glassy isomorphs starting at (a) $T=0.55$ and (b) $T=0.1$. Each figure shows 10 curves, where the density is increased by 1\% for each state point, giving a 9.4\% change in density overall. The insets show close-ups of (a) the first peak and (b) the second peak where very some small deviations are discernible.}
\end{figure}



The temperatures chosen for starting isomorphs are 0.55 and 0.1. For generating glassy isomorphs a configuration is drawn from the cooling run close to the desired temperature, and its density is used as the initial state for isomorphs. Due to fluctuations its density is not necessarily the same as mean density for the chosen temperature and pressure; similarly, when the doing NVT simulations in the glassy state the mean pressure is close to but not equal to the pressure of the cooling run. Table~\ref{tab:isomorph_data} shows thermodynamic information including the isomorph parameters $R$ and $\gamma$ for the different state point along each of the two isomorphs. By comparing the starting temperature of our high-temperature glass isomorph to the viscous liquid isomorph at the same density we estimate that it corresponds to temperature 0.42 at the usual density 1.2. At this temperature the Kob-Andersen mixture can be equilibrated as a liquid, but it requires substantial patience; at the strain rates we apply in our deformation runs, the system can be considered a glassy solid: According to Chattoraj {\em et al.}\cite{Chattoraj/Caroli/Lemaitre:2011}, particle displacements become driven more by strain than thermal motions once the strain rate exceeds $10^{-2}/\tau_\alpha$. Since our lowest strain rate is of order 10$^{-5}$ and $\tau_\alpha$ certainly exceeds 10$^3$, this criterion is met, and therefore we can speak of deformation of a glassy amorphous solid, at least regarding steady state dynamics. Our second isomorph, starting at the lower temperature 0.1, gives a system deep in the glassy state for which virtually no spontaneous relaxation is expected on conceivable simulation time scales. From Table~\ref{tab:isomorph_data} we see that the $R$-values for the lower temperature isomorph are somewhat lower than for the high-temperature isomorph, staying between 0.8 and 0.9; one might therefore expect poorer collapse of curves but we will see that this is not the case for our data.

Starting with the glassy states taken from the cooling run as mentioned above, we ran NVT simulations, then increased the density in steps of 1\%, while adjusting the temperature based on the observed value of $\gamma$, according to

\begin{equation}
T_{n+1} = T_n (1 + \gamma_n (\rho_{n+1}-\rho_n)/\rho_n)
\end{equation}

This procedure constitutes integrating Eq.~\eqref{conf_adiabat_gamma} numerically using the Euler method, and when applied to systems in equilibrium, generates curves of constant excess entropy. In applying it here we essentially ignore possible complications from being out of equilibrium, assuming for example, that no significant aging occurs during the simulation. The number of time steps is 10$^7$, and the starting configuration for each state point is the final configuration from the previous state point. Figure~\ref{glass_isomorph} shows three isomorphs in the density-temperature phase diagram including one equilibrium liquid isomorph and the two glassy isomorphs. Fig.~\ref{glassy_rdf} shows a very good degree of collapse for the radial distribution function (RDF) when plotted as a function of the reduced pair distance $\tilde r = \rho^{1/3} r$. This is true even for the high-temperature isomorph which one might expect to show some (small) changes of structure due to aging\cite{Kob/Barrat:2000}.
  

\section{\label{sec:stress_strain}Shear deformation: Analysis of stress strain curves}

When below the glass transition temperature (defined according to the accessible time scales) the system does not undergo any interesting dynamics unless perturbed by some external force. Due to time scale restrictions in simulations it is easiest to apply a large mechanical deformation to drive the system into a flowing state. In particular, we have chosen to apply simple (Couette) shear at a fixed strain rate and study the stress strain curve. For shearing we use the SLLOD algorithm combined with Lees-Edwards boundary conditions\cite{Evans/Morriss:1984,Ladd:1984,Allen/Tildesley:1987}. When identifying isomorph-invariant properties it is important that the shear rate be specified in an isomorph invariant way; that is, the reduced-unit strain-rate  $\tilde{\dot\gamma} = \dot\gamma (T/m)^{-1/2}\rho^{-1/3}$  should be fixed when comparing flowing states at different density-temperature points on an isomorph\cite{Separdar/others:2013}. The full set of flowing states is therefore characterized by a triple $(\rho, T, \tilde{\dot\gamma})$. Since the physics is in principle invariant along an $ \rho,T$-isomorph at a given reduced strain rate, we have thus a two-dimensional phase diagram, where a state can be labeled by isomorph and reduced strain rate. This has been previously shown in the non-viscous regime for the Lennard-Jones fluid\cite{Separdar/others:2013}, but has not been tested below the glass transition before. In our simulations we choose a ``nominal strain rate'' of 10$^{-2}$, 10$^{-3}$, 10$^{-4}$, or 10$^{-5}$, and nominal time step of 0.004. By ``nominal'' time step and strain rate we mean the value in real units at the first point of each isomorph.
These values are scaled by the appropriate powers of the density- and temperature-ratios to keep the reduced-unit time step and strain rate fixed along the isomorphs. For all our deformation runs we simulated 10$^8$ MD steps, which for the above nominal strain rates give total strains of 4000, 400, 40 and 4, respectively. Chattoraj {\em et al.} found that total strains of up to 1300\% or even 2400\% (i.e., 24) were necessary for accurate statistics\cite{Chattoraj/Caroli/Lemaitre:2011}. This suggests that our runs are sufficiently long except possibly for the lowest strain rates.  Note that the strain itself is dimensionless and therefore does not need to be put into reduced units.

\begin{figure}
\includegraphics[width=0.5\textwidth]{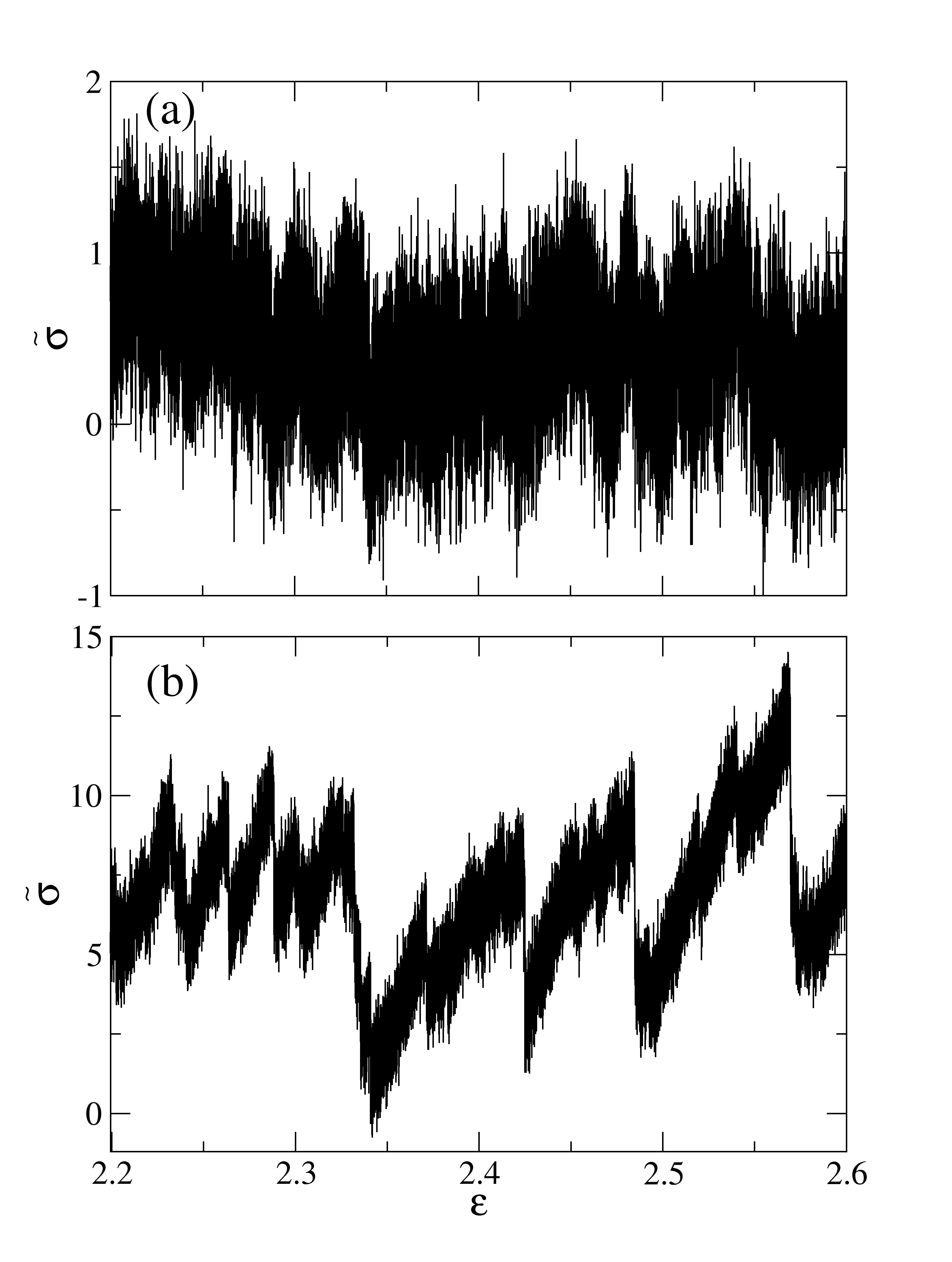}
\caption{\label{stress_strain_closeup} (a) Section of stress-strain curve for lowest-density state point on the higher-temperature isomorph ($\rho=1.265, T=0.550$) at the lowest nominal strain rate $10^{-5}$. (b) Section of stress-strain curve for a state point on the lower temperature isomorph ($\rho=1.324, T=0.100$) at the lowest strain rate $10^{-5}$. The abrupt  drops can be identified with avalanches of plastic activity. The difference in vertical scale between (a) and (b) can be attributed partly to the definition of reduced units for stress.}
\end{figure}

In principle isomorph theory predicts the whole stress strain curve to be invariant along isomorphs when stress is given in reduced units $\tilde\sigma=\sigma/\rho k_BT$. In the small systems typically studied in simulations, and particularly at low temperatures and strain rates, however, stress strain curves in the glassy regime exhibit extremely intermittent behavior\cite{Malandro/Lacks:1998,Maloney/Lemaitre:2004,Maloney/Lemaitre:2006,Bailey/others:2007} which is sensitive to initial conditions and other sources of randomness. Examples of this can be found in Fig.~\ref{stress_strain_closeup}.  Therefore a collapse of the actual stress-strain curves cannot be expected, except perhaps the initial part which covers the elastic regime and the transition to a flowing state. Instead we choose to study the statistical properties of the flowing state, in particular the steady state region where properties become time-independent, apart from fluctuations. We consider the steady state to have been reached after a strain of 50\%.

\begin{figure}
  \includegraphics[width=0.5\textwidth]{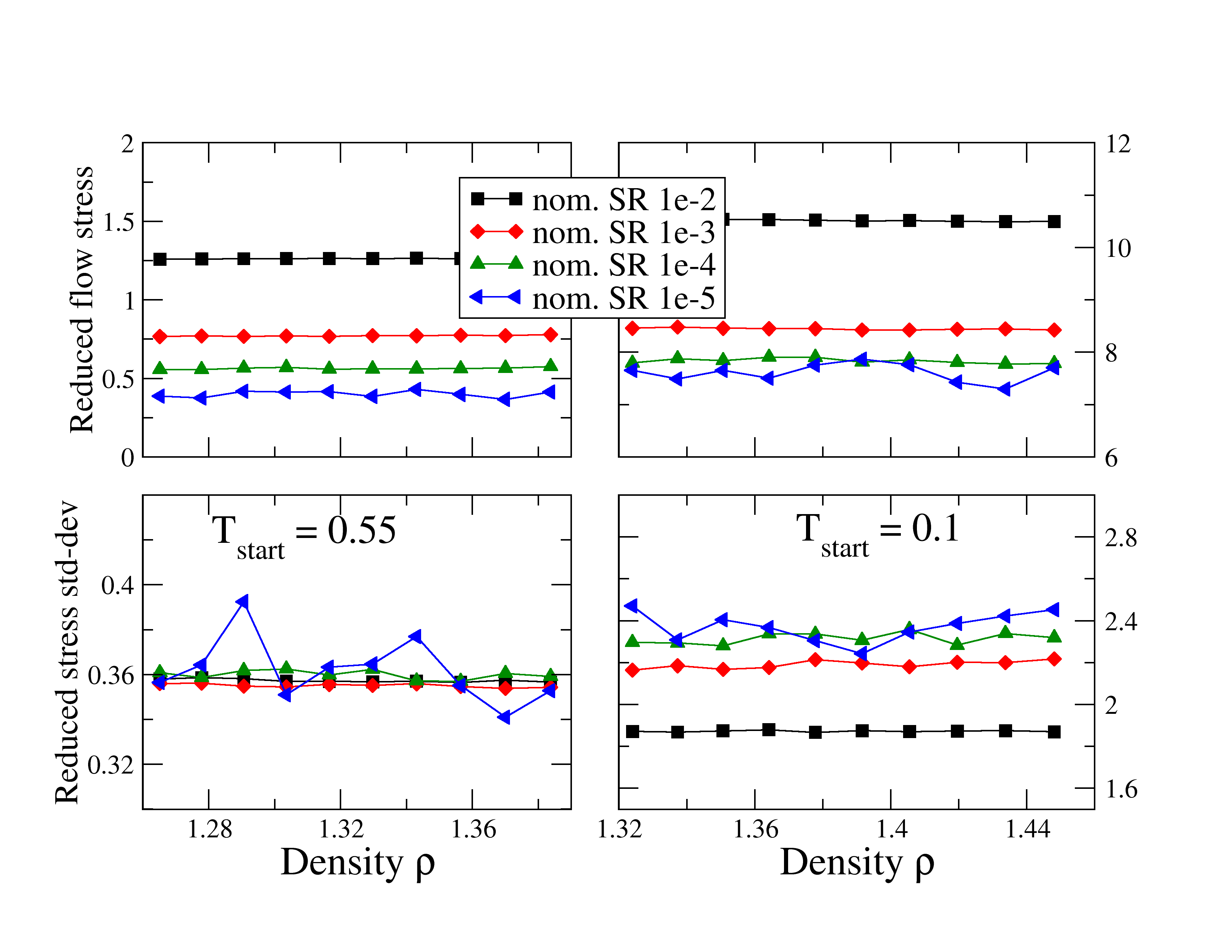}
  \caption{\label{stress_stats}Flow stress and standard deviation during steady-state regime as a function of density along an isomorph for different strain rates. The temperature at the starting density is 0.55 and the strain rate there is 10$^{-4}$; for other state points the reduced unit strain rate is the same.}
\end{figure}

The most basic statistical measures that can be extracted from the stress strain curve are the mean value of the stress (the flow stress) and the standard deviation. Figure~\ref{stress_stats} shows these quantities plotted in reduced units along the two isomorphs studied, with different nominal strain rates. The curves are consistent with being all flat within the statistical error; note that the latter is rather large for the standard deviation at the lowest strain rates (where, as we noted above, the total strain is probably not sufficiently large). This demonstrates isomorph invariance, the focus of this work. We can also comment briefly on the strain rate and isomorph dependence. The dependence of flow stress on strain rate is relatively weak given three orders of magnitude variation in the latter. Equivalently the shear viscosity varies by some orders of magnitude, indicating we are in a strongly non-Newtonian (shear-thinning) regime, as expected for glassy systems\cite{Berthier/Barrat:2002,Voigtmann:2014}. Comparing the two isomorphs, the reduced flow stress is a almost factor of ten smaller at the high temperature isomorph compared to the low temperature one, partly reflecting its proximity to the supercooled liquid state, but to some extent also an artifact of our choice of reduced units; see Sec.~\ref{discussion} below for discussion of alternative choices. Interestingly, for the high temperature isomorph the fluctuations of the stress are independent of strain rate (as well as being invariant along the isomorph). This must mean the fluctuations here are essentially thermal in origin, despite the rheology being clearly glassy in this regime (as determined from the strain rate dependence of the flow stress).

\begin{figure}
  \includegraphics[width=0.5\textwidth]{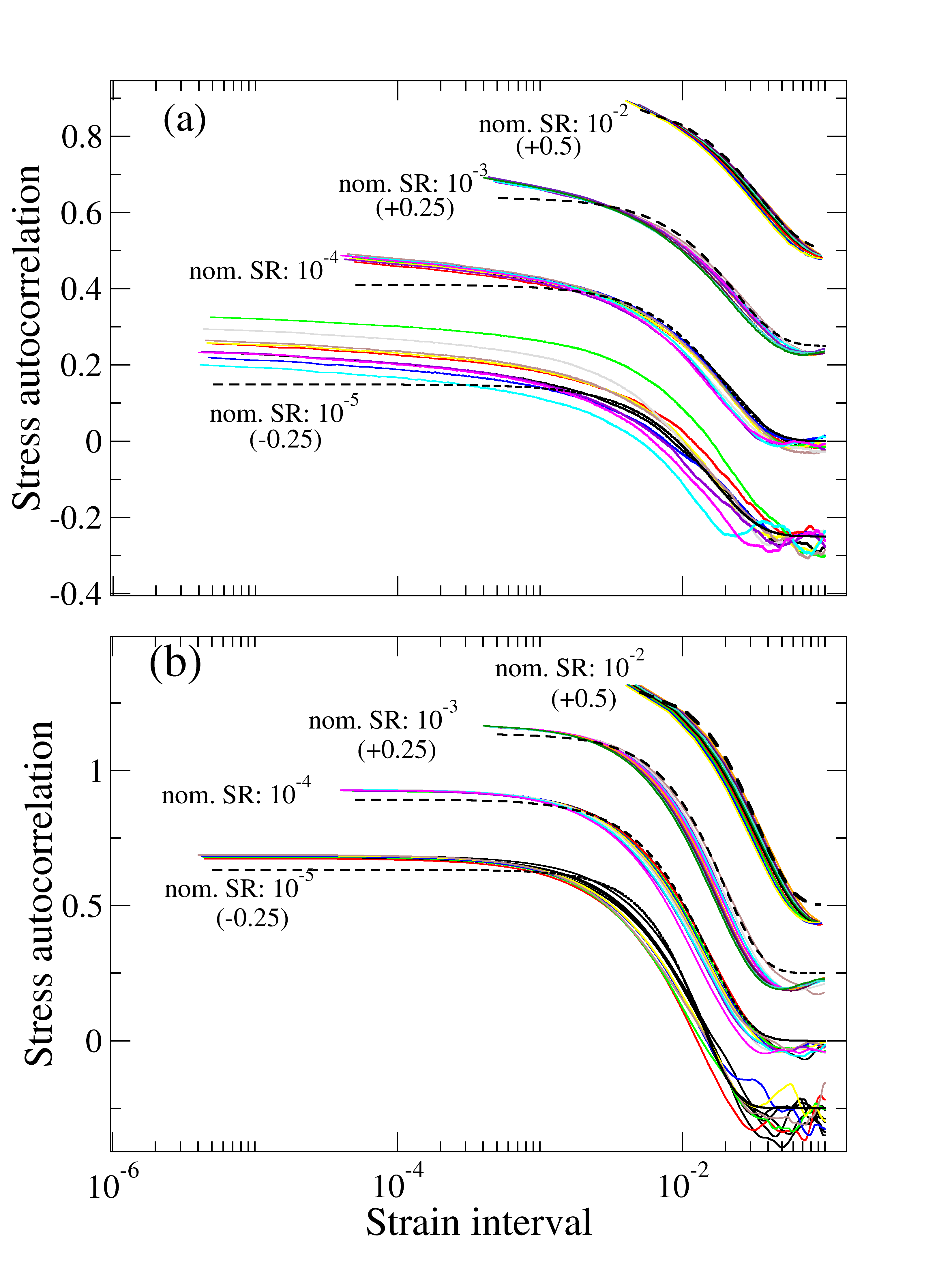}
  \caption{\label{sts_autocorr} Normalized shear stress autocorrelation functions along the high (a) and low (b) temperature isomorphs for three different strain rates. Curves have been shifted for clarity. The dashed lines indicated fits using a compressed exponential function for the first curve in each set (lowest density and temperature).}
\end{figure}

To investigate the dynamical correlations present in the stress strain curves, and check these for isomorph invariance, we consider the autocorrelation function of the shear stress, plotted as a function of strain interval. Fig.~\ref{sts_autocorr} shows the results. The collapse is not as good as we have seen in the flow stress. While the curves are somewhat noisy, inspection of the curves shows a trend whereby the de-correlation moves to lower strain intervals as density increases along the isomorph. To illustrate this more clearly we fit the autocorrelation curves to a compressed exponential,
  
\begin{figure}
  \includegraphics[width=0.5\textwidth]{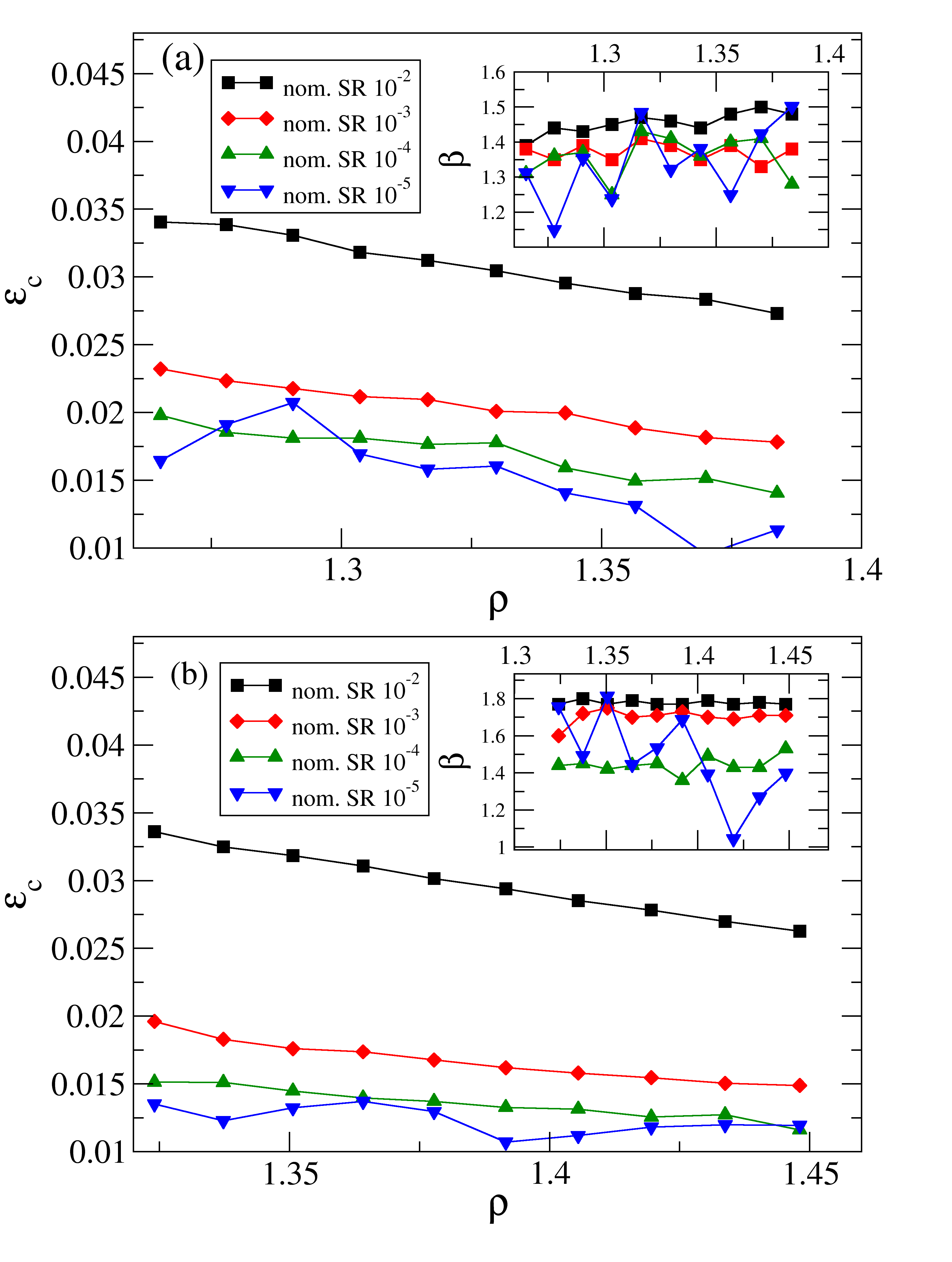}
\caption{\label{compressed_exp_fits} Fits of shear stress autocorrelation to Eq.~\eqref{compressed_exp} shown as functions of density along the high (a) and low temperature (b) isomorphs. The characteristic strain over which decays occurs, $\varepsilon_c$, decreases approximately linearly as density increases.}
\end{figure}

\begin{equation}\label{compressed_exp}
C(\Delta\varepsilon) = A\exp(-(\Delta\varepsilon/\varepsilon_c)^\beta),
\end{equation}
where $\beta$ is greater than unity. For $\beta<1$ this function is known as a stretched exponential, typically used to fit time-dependent relaxation correlation curves in the dynamics of supercooled liquid. The characteristic strain $\varepsilon_c$ corresponds to the relaxation time $\tau$ in time-dependent correlation functions, indicating roughly the strain interval after which a stress fluctuation has decayed away. As shown in Fig.~\ref{sts_autocorr}, the compressed exponential can fit the main part of the decay reasonably well, but not the initial slow decay, with values of the characteristic strain $\varepsilon_c$ falling in the range 0.01-0.035, and values of the compression exponent $\beta$ in the range 1.3-1.5. Along the isomorphs, $\varepsilon_c$ decreases approximately linearly as density increases, in a similar manner for both isomorphs, while  $\beta$ increases slightly for the high temperature isomorph but shows little variation on the low temperature isomorph. Comparing different strain rates, both $\varepsilon_c$ and $\beta$ decrease as strain rate decreases, although for $\beta$ the the effect is weak compared to noise. Further investigation with longer runs will be necessary to determine if the apparent variation of $\varepsilon_c$ is an artifact of insufficiently long runs, a sign of an imperfect procedure for generating isomorphs, or a genuine limit of isomorph invariance (which is never exact). We note also that there seems to be a systematic undershoot to negative correlation, after  most of the stress has de-correlated. This could tentatively be interpreted as a sign of avalanche-type dynamics; see below.


As a further type of statistical analysis of the stress strain curves we attempt to infer something about the microscopic processes by considering the distributions of stress changes $\Delta\sigma$ over a given interval of strain $\Delta\varepsilon$. Unlike the case of athermal, infinitely slow driving that has been studied by several authors\cite{Malandro/Lacks:1998,Maloney/Lemaitre:2004,Maloney/Lemaitre:2006,Bailey/others:2007,Lerner/Procaccia:2009b,Lerner/Procaccia:2010,Lerner/Bailey/Dyre:2014}, it is not possible to unambiguously identify single flow events or so-called ``avalanches'', since thermal fluctuations tend to merge them together. Lema\^itre {\em et al.} have, however, shown that the dynamics of a glassy system can still be understood in terms of avalanche-type behavior at relatively high temperatures\cite{Chattoraj/Caroli/Lemaitre:2010,Chattoraj/Caroli/Lemaitre:2011}. Moreover, visual inspection of the stress-stress curves for lower strain rates and temperatures shows drops in the stress reminiscent of avalanche behavior, see Fig.~\ref{stress_strain_closeup}(b). The distribution of stress changes over a given strain interval can be used to identify signatures of avalanche behavior without having to identify precisely when avalanches occur.

\begin{figure*}
  \includegraphics[width=\textwidth]{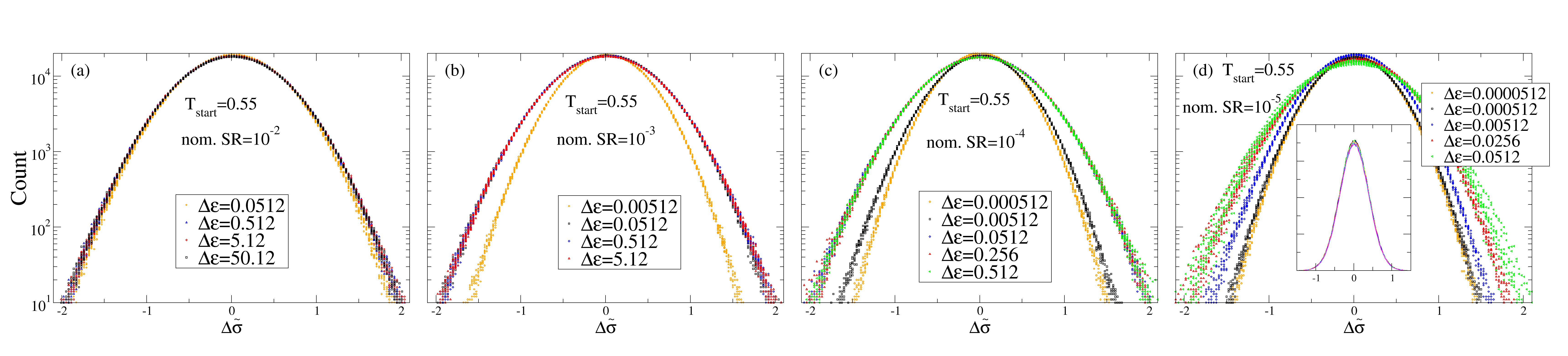}
\caption{\label{hist_delta_stress_highT} Histograms of stress changes of intervals as indicated for the high-temperature isomorph for different strain rates. The distributions are essentially Gaussian for all strain rates and strain intervals $\Delta\varepsilon$, and their widths are relative insensitive to $\Delta\varepsilon$ even at the lowest strain rates, indicating that most of the fluctuations are thermal rather than strain-driven. In the main figures distributions for different isomorphs, for a given $\Delta\varepsilon$, have been plotted in the same color; the inset in (d) shows the distributions for $\Delta\varepsilon$=0.000512 with the different members of the isomorph colored differently, on a linear scale, to emphasize the isomorph invariance of the distributions. }
\end{figure*}

\begin{figure*}
  \includegraphics[width=\textwidth]{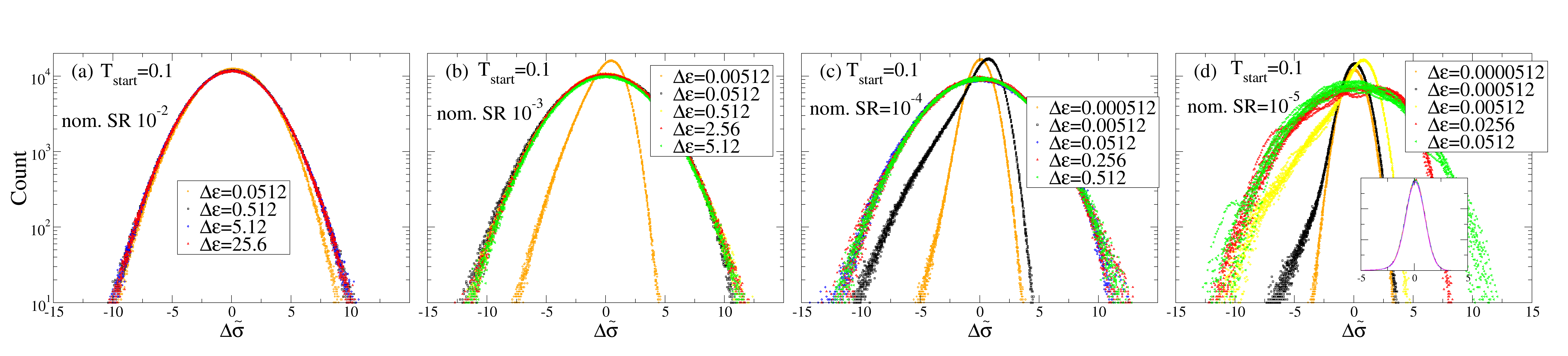}
\caption{\label{hist_delta_stress_lowT}Histograms of stress changes of intervals as indicated for the low-temperature isomorph for different strain rates. They are Gaussian for the largest strain intervals $\Delta\varepsilon$ as well as for the shortest $\Delta\varepsilon$ at the slowest strain rate, where the contribution of strain to the fluctuations is negligible compared to the thermal contribution. For larger $\Delta\varepsilon$ at the slowest strain rate an exponential tail on the negative side is a clear indication of plastic events organizing into avalanches. For even larger $\Delta\varepsilon$, and at the larger strain rates, mixing of thermal and mechanic noise, and multiple avalanches lead to more disorganized histograms. The inset of (d) shows on a linear scale distributions of the second smallest strain interval with the different members of the isomorph represented with different colors to emphasize the invariance.}
\end{figure*}

Figures \ref{hist_delta_stress_highT} and \ref{hist_delta_stress_lowT} show histograms of the reduced unit stress changes, $\Delta\tilde\sigma=\Delta\sigma/\rho k_BT$,  for different strain intervals $\Delta\varepsilon$ and different strain rates, from simulations on the high and low temperature isomorphs, respectively. Curves of the same color represent data from different state points on the isomorph and the near collapse shows that the statistics as probed by these histograms is isomorph invariant to a high degree. This can be seen more explicitly in the insets of Fig.~\ref{hist_delta_stress_highT}(d) and \ref{hist_delta_stress_lowT}(d) where the distributions for the different members of the corresponding isomorph are shown in different colors, for one particular strain interval. Having demonstrated isomorph invariance it is interesting to note some of the other features of these data. One feature common to both isomorphs and all strain rates, is that for sufficiently large $\Delta\varepsilon$ -- over 0.05 -- the histograms converge to a Gaussian  whose variance is twice that of the stress fluctuations (mostly within 1\%; 10\% for the slowest two strain rates at the lower temperature isomorph, where the statistical errors are larger). This is expected since our analysis of the autocorrelation indicates that correlations vanish by strain 0.05 in all cases, see Fig.~\ref{sts_autocorr} (the characteristic strain interval for decay is between 0.015 and 0.035, with the functions essentially reaching zero by 0.05). For smaller intervals $\Delta\varepsilon$ the distribution is generally narrower and reflects contributions to stress fluctuations from the mechanical driving as well as from thermal fluctuations. As noted above these cannot be necessarily separated, but a reasonably clear picture emerges from considering the dependence on isomorph, strain rate, and $\Delta\varepsilon$.

Focusing first on the high-temperature isomorph, Fig.\ref{hist_delta_stress_highT} (a)-(d) shows stress change histograms for strain rates 10$^{-2}$, 10$^{-3}$, 10$^{-4}$ and 10$^{-5}$ respectively. For all strain rates the distribution converges to the same Gaussian at large intervals $\Delta\varepsilon$. This is consistent with the lower right panel of Fig.~\ref{stress_stats}, which showed that the fluctuations of the stress strain curve are independent of strain rate (as well as being isomorph invariant) in the high-temperature case -- a sign that the fluctuations are dominated by thermal noise in this regime. For the high strain rate the shortest interval is already 0.05 so we see no dependence on interval here. Some dependence on strain interval can be seen at low strain rate where the width of the distribution appears to converge to a lower value in the limit of small strain intervals. The time scale for the shortest interval is of order 5 Lennard-Jones units (at the lowest-density point on the isomorph) which should be still somewhat longer than the vibrational time scale, therefore this apparent limit presumably represents the full thermal contribution to the fluctuations for an undeformed glassy system. The increased width at high intervals can therefore be interpreted as coming from the sampling of different glassy configurations due to deformation\footnote{Note that this would presumably also happen even without  any deformation by waiting long enough for liquid dynamics to become relevant --- in that case time, rather than strain, becomes the relevant parameter.}.

Figure \ref{hist_delta_stress_lowT} shows histograms for the lower-temperature isomorph and the same nominal strain rates as Fig.~\ref{hist_delta_stress_highT}. More interesting behavior is apparent at these low temperatures, particularly at the lowest strain rates for example (nominal) 10$^{-5}$: for the shortest intervals we see a Gaussian, representing purely thermal fluctuations which are small at this temperature.  In other words, for a strain interval of 0.00005 the stress change due to driving is hidden by the thermal fluctuations. As discussed above we see a Gaussian at the largest intervals where all correlations have decayed. For intermediate strain intervals, however, a marked deviation from Gaussian behavior appear in the form of a roughly exponential tail on the negative side. This is a clear indication of avalanches: correlated aggregations of multiple microscopic flow events which release the stress, giving large negative stress changes as studied in the quasi-static case\cite{Malandro/Lacks:1998,Maloney/Lemaitre:2004,Maloney/Lemaitre:2006,Bailey/others:2007,Lerner/Procaccia:2009b,Lerner/Procaccia:2010,Lerner/Bailey/Dyre:2014}.

An analysis somewhat similar to ours was carried by Rottler and Robbins\cite{Rottler/Robbins:2003}, who also found exponential tails at low temperature and strain rate. Note that since we consider a steady state situation, the mean of the stress changes must be zero, implying that the main Gaussian is shifted slightly to positive values. We have checked this by fitting the Gaussian part (not shown). The positive mean of the non-avalanche fluctuations corresponds to elastic loading which is then released by the avalanches. In the limit of zero temperature and then infinitely slow deformation\cite{Maloney/Lemaitre:2006} the narrow Gaussian seen at short intervals would converge to a delta-function at a small positive value (the shear modulus times the strain interval).

\begin{figure}
  \includegraphics[width=0.5\textwidth]{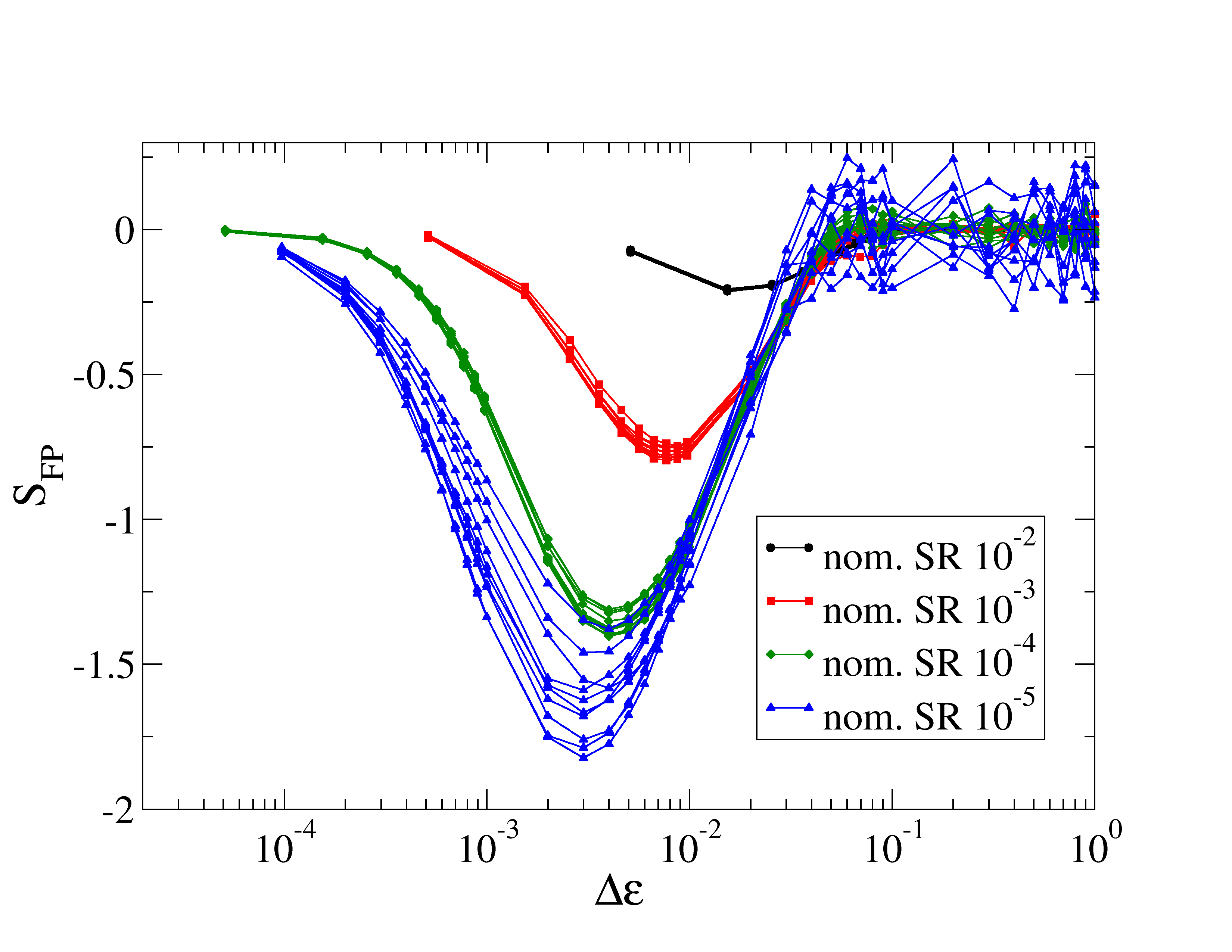}
  \caption{\label{skewness} Fisher-Pearson skewness $S_{FP}$ of (reduced) stress drop distributions as a function of strain interval for different strain rates for the low-temperature isomorph. Different curves of the same color correspond to different points on the isomorph.}
\end{figure}

\begin{figure}
  \includegraphics[width=0.5\textwidth]{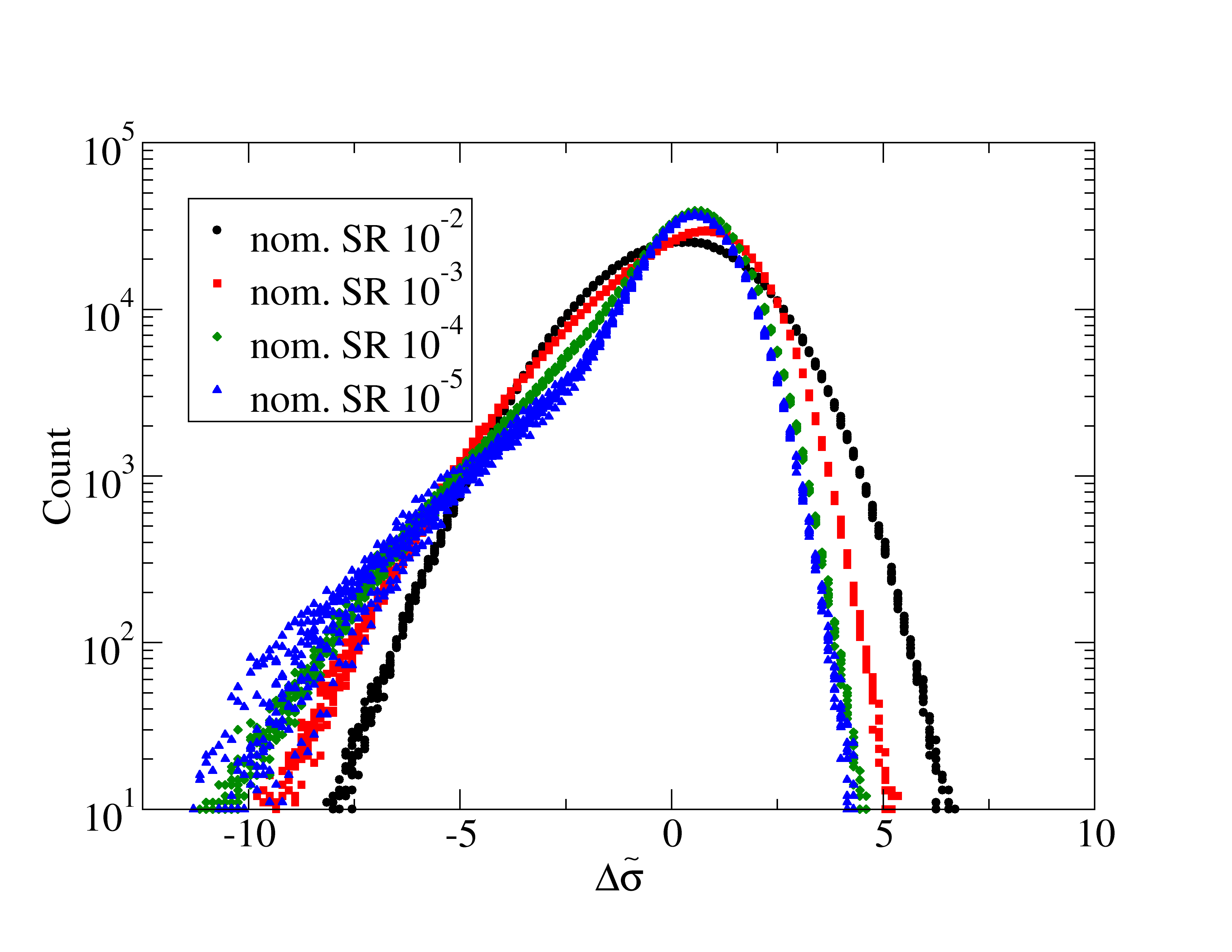}
\caption{\label{histograms_max_skewness} Histograms of (reduced) stress changes over strain intervals $\varepsilon_s$ chosen to minimize skewness for each strain rate, on the low temperature isomorph. Data for different points on the isomorph are plotted in the same color for each nominal strain rate. In order of decreasing strain the  minimum-skew strain intervals, as judged by eye from Fig.~\ref{skewness}, are 0.02, 0.008, 0.004, 0.003. }
\end{figure}

The asymmetric deviations from Gaussianity can be quantified by the Fisher-Pearson coefficient of skewness, based on the third moment of the distribution scaled by the cube of the standard deviation:
\begin{equation}\label{eq:SFP}
S_{FP} = \frac{m_3}{m_2^{3/2}},
\end{equation}
where
\begin{equation}
m_i = \frac{1}{N} \sum_{n=1}^N (x_n-\bar x)^i
\end{equation}
where $\bar x$ is the sample mean. 
Fig.~\ref{skewness} shows $S_{FP}$ as a function of strain interval for four different strain rates for the low temperature isomorph. Different curves with the same color come from different members of the isomorph for a given strain rate. The skewness vanishes for short and long strain intervals where, as discussed above, the distributions become Gaussian. The variations between the distributions for a given strain rate are not systematic, and thus presumably reflect statistical uncertainty. The variation is relatively small and thus consistent with this measure of the dynamics being isomorph invariant (this follows of course also from the good collapse of the distributions in Figs.~\ref{hist_delta_stress_highT} and \ref{hist_delta_stress_lowT}). The minimum (most negative) value of the skewness parameter identifies a strain interval $\varepsilon_s$ at which the deviation from Gaussianity is most pronounced. Histograms for this strain interval are plotted in Fig.~\ref{histograms_max_skewness} for the low temperature isomorph and different strain rates, with the values of  $\varepsilon_s$ given. These values are a factor of 2-3 smaller than the characteristic strain intervals identified from the autocorrelation functions, see Fig.~\ref{compressed_exp_fits}. For the lowest strain rate it is an order of magnitude {\em larger} than the strain interval at which the exponential tail indicating avalanche behavior is seen, 5$\times 10^{-4}$ (see Fig.~\ref{hist_delta_stress_lowT}). Denoting the latter $\varepsilon_a$ (where $a$ denotes avalanche), we can tentatively identify, in the low temperature, low strain-rate limit at least, a broad hierarchy of strain scales which characterize different physical processes:

\begin{enumerate}
\item The smallest strain scales where stress fluctuations are purely thermal/vibrational.
\item The ``avalanche'' strain $\varepsilon_a$ over which stress changes show signs of correlated, avalanche-type behavior, of order 5$\times 10^{-4}$.
\item The strain over which stress change distributions deviate most from Gaussianity, $\varepsilon_s$, an order of magnitude larger than $\varepsilon_a$. Here the exponential tails of the avalanches, and the changes due elastic loading between them, merge to make a broader distribution, but signs of correlation remain.
\item The characteristic strain $\varepsilon_c$ identified via the stress autocorrelation function.  $\varepsilon_c$ is of order $2 \times 10^{-2}$ which is a small factor (2-3) larger than $\varepsilon_s$.
\item Finally there the strain interval around $5 \times 10^{-2}$ beyond which all correlation has vanished (though this is not physically independent from $\varepsilon_c$; rather it represents where the autocorrelation function is small compared to $1/e$).
\end{enumerate}



\section{\label{sec:particle_dymamics}Particle dynamics under shear: transverse diffusivity}


\begin{figure}
  \includegraphics[width=0.5\textwidth]{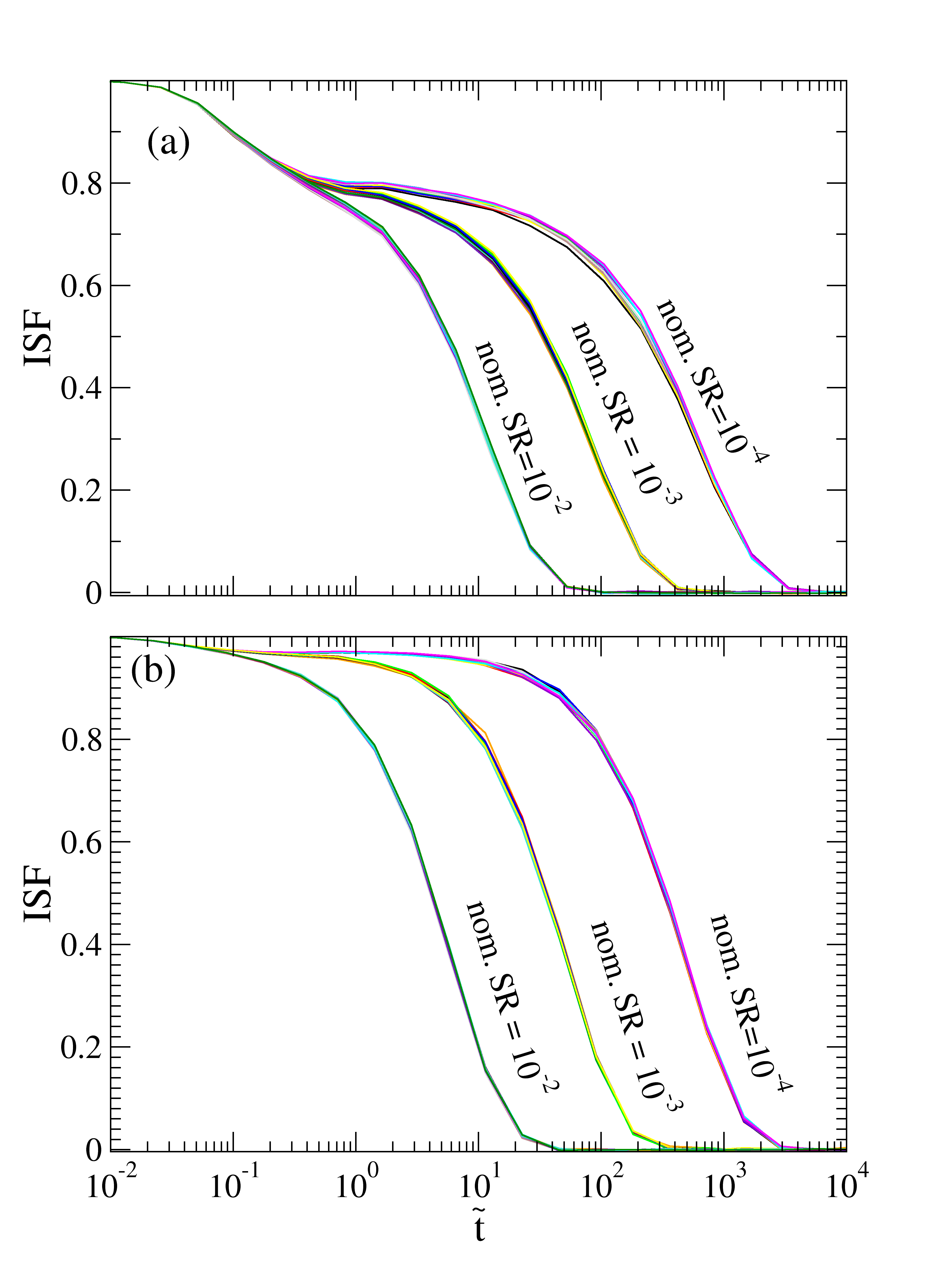}
  \caption{\label{isf}Self-part of the intermediate scattering function for larger (A) particles based on particle displacements transverse to the shearing direction for (a) the high temperature isomorph and (b) the low temperature isomorph, for different strain rates.}
\end{figure}

As an alternative probe of dynamical processes under steady state shearing we consider also the particle displacements. Accounting for the contribution to a particle's displacement in the shearing direction when using Lees-Edwards boundary conditions is non-trivial\cite{Lemaitre/Caroli:2007}, so we consider only the components of a particles displacement transverse to the shearing direction. Based on these displacements we compute the self-intermediate scattering function (ISF) and the mean squared displacement (MSD). For the ISF one must choose a wavenumber $q$, which as is conventional we choose to be near the first peak in the structure factor $S(q)$. This must be of course scaled according to $\rho^{1/3}$ along an isomorph, such that the reduced wavenumber $\tilde q \equiv q \rho^{-1/3}$ is constant (this is compatible with choosing $q$ to be near the first peak, as $S(q)$ is also invariant in reduced units)\cite{Gnan/others:2009}. We restrict to the larger (A) particles for brevity. The ISF is shown in Fig.~\ref{isf}.

For both isomorphs and all strain rates we find a good collapse, though slightly less so for the lowest strain rates. Fitting of the curves to a stretched exponential form (Eq.~\eqref{compressed_exp} where $\beta<1$ and with $\tau$ instead of $\varepsilon_c$), not shown, indicates at most a slight systematic variation in relaxation time $\tau$, suggesting the apparent failure to collapse perfectly is mostly due to statistical error. From the fits, for the high temperature isomorph we find near exponential behavior ($\beta\simeq 1$) for the highest strain rates and mildly stretched exponential behavior as the strain decreases ($\beta \simeq 0.85$ at the lowest strain rate). For the low temperature isomorph we find near exponential behavior for all strain rates. Stretched exponential behavior is typical of dynamics in the supercooled, highly viscous liquid. The vanishing of stretching (i.e. the near-exponential behavior) at low temperatures and slow shearing indicates that the nature of particle dynamics is different in this regime. Exponential behavior of the self-intermediate scattering function for the same system under shear in the limit of zero temperature was also reported some years ago by Berthier and Barrat\cite{Berthier/Barrat:2002}.

\begin{figure}
  \includegraphics[width=0.5\textwidth]{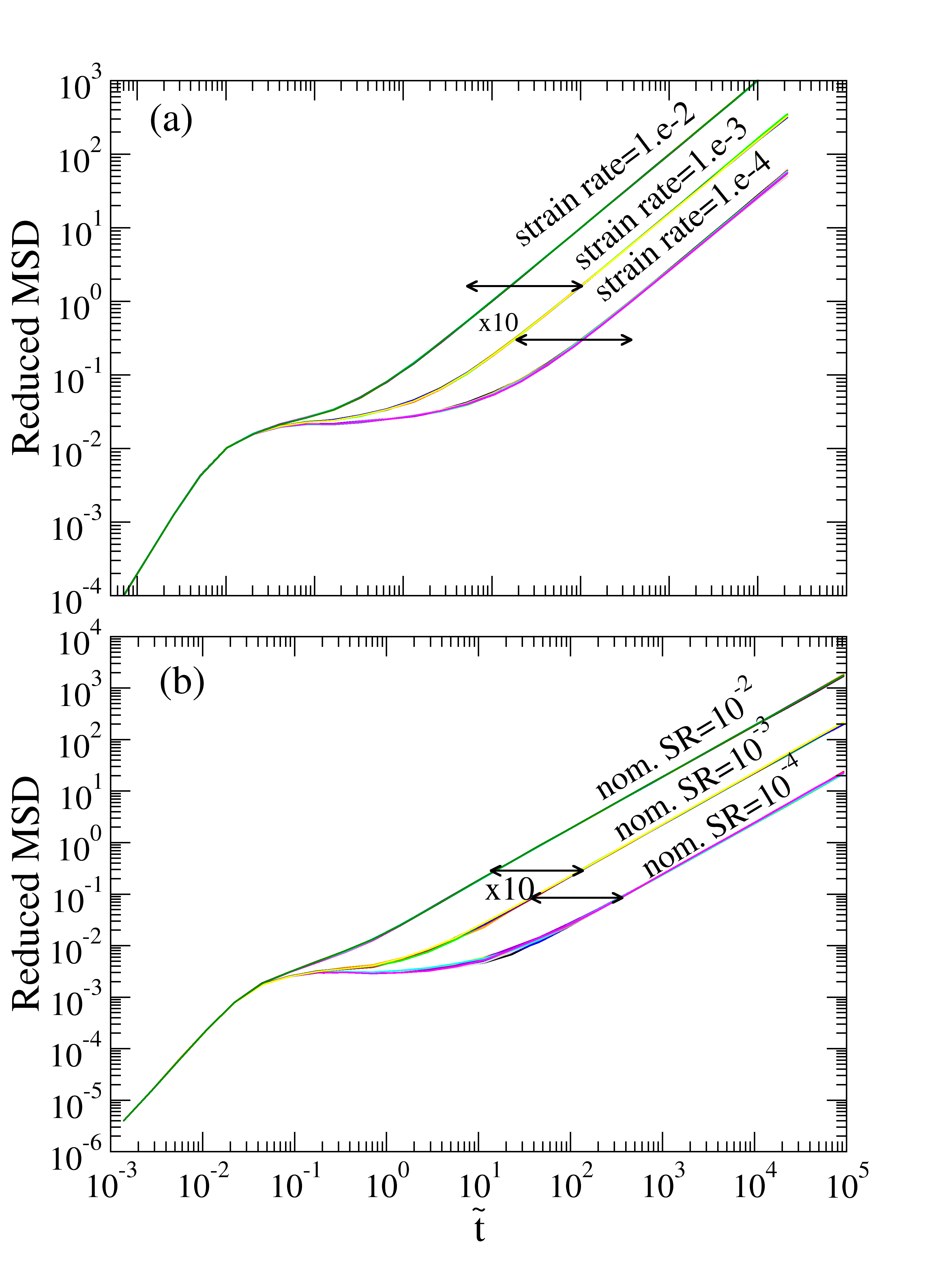}
  \caption{\label{msd} Mean squared transverse displacement plotted in reduced units for (a) high temperature isomorph and (b) low temperature isomorph. The horizontal arrows indicate a factor of ten in the time axis, and can be used to judge by what factor the curves can be shifted onto each other in time.}
\end{figure}

\begin{figure}
  \includegraphics[width=0.5\textwidth]{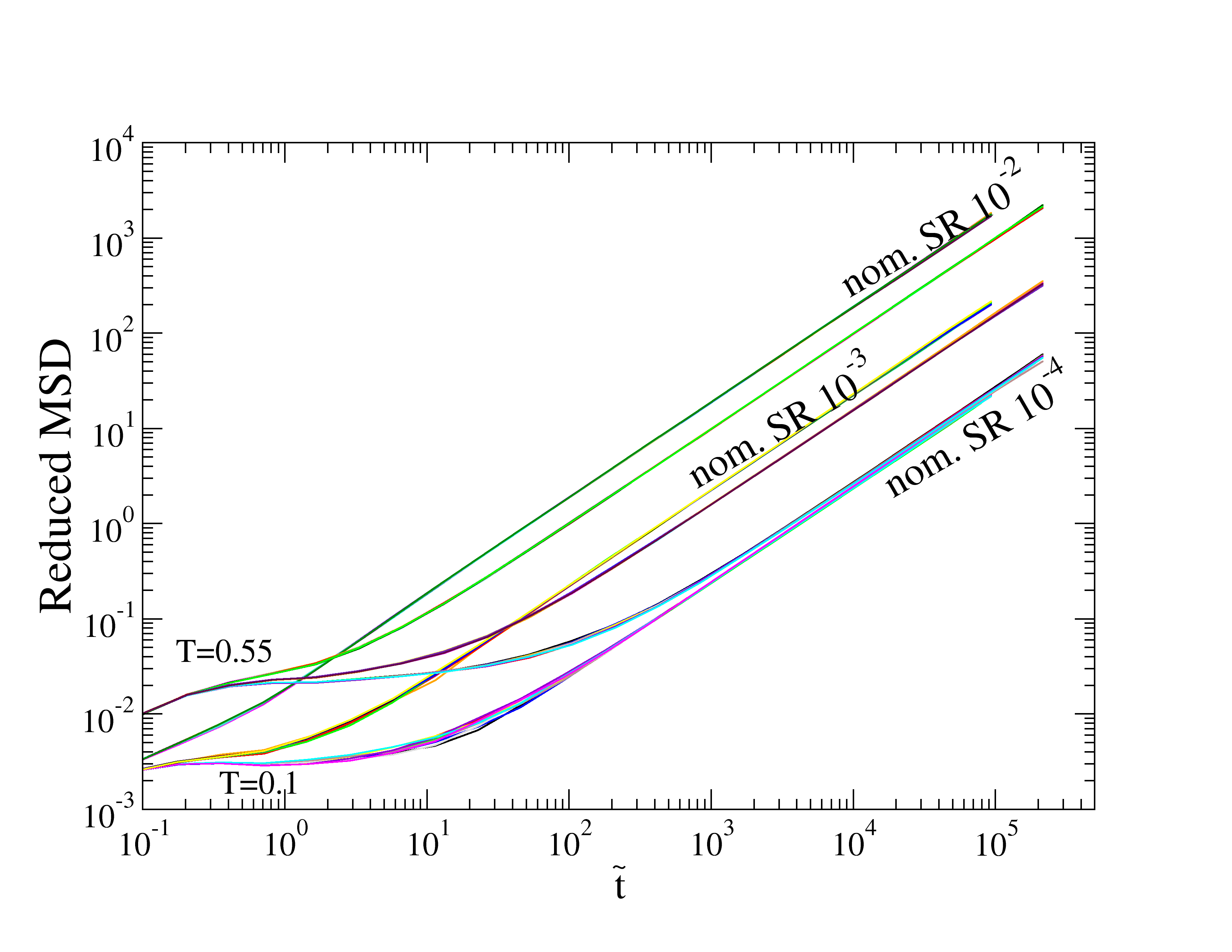}
\caption{\label{show_all_msd} The MSD curves from Fig.~\ref{msd} (a) and (b) plotted together, though without the short-time parts. At low strain rates the MSD appears to become independent of isomorph, as well as which point on the isomorph. The definition of reduced units means that the curves for the different isomorphs are plotted in terms of different time scales, so caution is required when drawing conclusions from the apparent collapse.}
\end{figure}

Plots of the mean squared transverse displacement in reduced units are shown in Fig.~\eqref{msd}. The form of the curves is reminiscent of what is seen for equilibrium viscous liquids: a ballistic regime at short times (where the slope is 2), a plateau of varying extent, followed by a transition to diffusive behavior (slope 1). The collapse is good in all cases, although again some deviations are apparent for the lowest strain rates, particularly around the crossover to diffusive behavior. Superficially not much difference can be seen between the low and high temperature isomorphs, but upon closer examination one can see some physically relevant differences. On the higher temperature isomorph thermal motion is greater, thus the height of the plateau (in units of the interparticle spacing) is larger. More interestingly, in the low temperature case, the diffusivity curves are essentially a factor of ten apart in the time axis, corresponding to the factor ten change in strain rate, while for the high temperature case the diffusivity curves are shifted by a factor smaller than 10 in the time axis. The interpretation is that thermal activation plays a noticeable role in particle diffusion in the high temperature case, but almost no role in the low-temperature case. In the latter the diffusive motion is determined entirely by the strain rate at the lowest temperatures. Fig.~\ref{show_all_msd} shows the long-time parts of the MSD for both isomorphs. In this plot the difference at long times between the two isomorphs appears minimal -- the MSD is determined much more by the strain rate than by which isomorph is considered (and almost not at all by which point on the isomorph, which is the essence of isomorph invariance). It must be noted, however, that direct numerical comparison of the MSD curves at different temperatures (isomorphs) for the same nominal strain rate can be difficult to interpret due to the use of reduced units, thus it appears that at nominal strain rate $10^{-2}$ that the diffusivity, counterintuitively, is greater on the lower temperature isomorph. Recall though, that this is in reduced units, i.e. with respect to a time scale defined by the thermal velocity. A meaningful comparison would first of all involve identical reduced (rather than nominal) strain rates -- the reduced strain rates for the low temperature isomorph are a factor of 2.3 higher than the corresponding ones for the high temperature isomorph. Second, there is a further complication already alluded to, which will be discussed further below, namely that the definition of reduced units is not unique, and a different choice could in principle be more relevant, and elucidate the physics better, in the limit of low temperatures. We emphasize that the most important result in this section is the near perfect collapse of the MSD for different state points along a given isomorph (and given reduced strain rate), when reduced units are used.


Lema\^itre and coworkers have studied over several papers the effect of finite temperatures and strain rates on avalanche dynamics\cite{Lemaitre/Caroli:2009, Chattoraj/Caroli/Lemaitre:2010, Chattoraj/Caroli/Lemaitre:2011}. They found that studying transverse particle diffusivity is useful for disentangling the effects of strain and temperature. In particular Chattoraj {\em et al.}\cite{Chattoraj/Caroli/Lemaitre:2011}  used the transverse diffusivity $D$ determined from the long-time limit of the MSD curves, and its strain-normalized analog $D/\dot\gamma$. Their Fig.~5 shows nicely the crossover from strain dominated to temperature dominated diffusion. As they point out, the strain-normalized diffusivity is the more relevant one in the strain-driven regime (low temperatures and strain rates) while normal diffusivity is relevant at high temperatures. Moreover they show that the crossover strain-rate as a function of temperature tracks more or less the inverse relaxation time: strain begins to have a pronounced effect on particle diffusion once the strain per relaxation time exceeds an amount of order $10^{-3}-10^{-2}$. Our results are consistent with theirs in terms of the interplay of strain-driven and thermal contributions to particle motion. They did not consider density as a parameter, but our results show that it can be simply accounted through isomorph invariance, and by remembering that by at ``high temperature'' is really meant ``on high-temperature isomorphs''.




\section{\label{discussion}Discussion}

\subsection{Implications for theories for flow stress}

Several authors have studied the dependence of flow stress of simulated amorphous solids below the glass transition on thermodynamic parameters such as density, temperature and strain rate and system size\cite{Rottler/Robbins:2003, Lerner/Procaccia:2009a,Karmakar/others:2010,Hentschel/others:2010,Chattoraj/Caroli/Lemaitre:2011}. System size becomes relevant for the flow stress at the lowest temperatures where deformation occurs through avalanches\cite{Lemaitre/Caroli:2009}. Some of these works have attempted to determine theoretical expressions or scaling forms to account for size-, temperature- and strain rate-dependence of the rheology of amorphous solids\cite{Karmakar/others:2010, Chattoraj/Caroli/Lemaitre:2011,Hentschel/others:2010}, while only few have included density as a variable\cite{Lerner/Procaccia:2009a}. One of the crucial implications of the existence of isomorphs is that it doesn't make sense to think of temperature dependence in isolation from density dependence. We therefore hope that future theoretical work on the rheology of amorphous solids will take this into account. To illustrate this point we consider the expression developed in Ref.~\onlinecite{Chattoraj/Caroli/Lemaitre:2011} for the flow stress as a function of $T$ and $\dot\gamma$,

\begin{equation}\label{Chattoraj_flow_stress_v1}
\sigma(\dot\gamma, T) = A_0 + A_1\sqrt{\dot\gamma} - A_2T^{2/3}[\ln(A_3T^{5/6}/\dot\gamma)]^{2/3}
  \end{equation}
  where $A_1$, $A_2$, $A_3$, $A_4$ are constants. The form of the expression and the interpretation of the constants were derived through a combination of the theoretical considerations and fitting to data from 2D simulations. We make no claims regarding its validity for 3D situations, but rather wish to illustrate how this expression can be made isomorph invariant. We assume 3D in the sense that $\rho$ has units of inverse length cubed. To include density dependence we must allow the $A_i$ to be functions of density whose functional form will be determined by isomorph theory. Using standard reduced units (an alternative will be discussed below), we re-write Eq.~\eqref{Chattoraj_flow_stress_v1} in terms of the reduced flow stress $\tilde\sigma\equiv\sigma/(\rho k_B T)$ and strain rate $\tilde{\dot\gamma}$:

\begin{multline}\label{Chattoraj_flow_stress_v2}
  \tilde\sigma(\dot\gamma, \rho, T) = \frac{A_0(\rho)}{\rho k_BT} + \frac{A_1(\rho)}{\rho k_B T}\sqrt{\tilde{\dot\gamma}} \rho^{1/6}\left(k_BT/m\right)^{1/4} \\
  - \frac{A_2(\rho)}{\rho k_BT} T^{2/3}[\ln(A_3(\rho)T^{5/6}\rho^{-1/3}(k_BT/m)^{-1/2}/\tilde{\dot\gamma})]^{2/3}
  \end{multline}
  
To proceed from here we recall from earlier work\cite{Ingebrigtsen/others:2012} that an isomorph in the density -- temperature plane may be written as $h(\rho)/k_BT=$constant, where the constant indexes the isomorph. The function $h(\rho)$ has not been used so far in the present work; it is been the basis of theoretical analysis connecting the shape of the isomorphs to the interatomic potential\cite{Ingebrigtsen/others:2012, Boehling/others:2012}. $h(\rho)$ describes the the way the potential energy surface depends on density and is sometimes called the {\em density scaling function}. The assumption that this depends only on density (and not on which isomorph one considers) is equivalent to assuming $\gamma$ depends essentially only on $\rho$. Indeed, $\gamma$ is then simply the logarithmic derivative of $h(\rho)$:

\begin{equation}
  \gamma(\rho) = \frac{d \ln h(\rho)}{d\ln \rho}
\end{equation}

The normalization of $h(\rho)$ is arbitrary, but it makes physical sense to assume that it has units of energy, since it describes the density scaling of the potential energy surface. If Eq.~\ref{Chattoraj_flow_stress_v2} is to hold in a system with good isomorphs then the individual terms  must be isomorph invariant. Specifically they must be writable as powers of the combination $h(\rho)/k_BT$. Taking the first as an example we find that $A_0=\tilde A_0 \rho h(\rho)$, where $\tilde A_0$ is a dimensionless constant (i.e., independent of both temperature and density). The full set is, as can be checked straightforwardly,

\begin{align}
  A_0(\rho) &= \tilde A_0 \rho h(\rho) \\
  A_1(\rho) &= \tilde A_1 \rho^{5/6} (h(\rho))^{3/4} \\
  A_2(\rho) &= \tilde A_2 \rho (h(\rho))^{1/3} \\
  A_3(\rho) &= \tilde A_3 \rho^{1/3} (h(\rho))^{-1/3}
\end{align}
Inserting these expressions into Eq.~\eqref{Chattoraj_flow_stress_v2} yields the following expression for the reduced flow stress as a function of the isomorph scaling combination $h(\rho)/k_BT$ and the reduced strain rate:

\begin{multline}\label{Chattoraj_flow_stress_v3}
  \tilde\sigma(\dot\gamma, \rho, T) = \tilde A_0 \frac{h(\rho)}{k_BT} + \tilde A_1\left(\frac{h(\rho)}{k_B T}\right)^{3/4}\sqrt{\tilde{\dot\gamma}} \\
  - \tilde A_2\left(\frac{h(\rho)}{k_BT}\right)^{1/3}[\ln(\tilde A_3\left(\frac{h(\rho)}{k_BT}\right)^{-1/3}/\tilde{\dot\gamma})]^{2/3}
\end{multline}
  This is an explicitly isomorph invariant theoretical expression for the flow stress as a function of $\rho$, $T$ and $\dot\gamma$, based on the original expression whose validity was determined (or assumed) for a particular density. However given that the original expression had a finite limit as $T\rightarrow0$, it seems problematic that the reduced stress therefore diverges as we consider isomorphs lower and lower in temperature. Therefore we must consider alternative definitions of reduced units when approaching zero temperature.

\subsection{Alternative reduced units}

Our definition of reduced units, apart from the length unit, is based on thermal motion; thus the energy scale is $e_0 = k_BT$, the velocity scale is $v_0=(k_BT/m)^{1/2}$ and the time scale is the time for a particle with such a constant velocity $v_0$ to cross the interparticle spacing, $t_0=\rho^{-1/3} (k_BT/m)^{-1/2}$. This choice has the advantage of using only macroscopic parameters; apart from the particle mass, no knowledge about the system under consideration (its Hamiltonian, phase diagram or isomorphs) is needed. But as noted above this definition becomes problematic as temperature approaches zero --- it is natural at finite temperature but not in the limit of zero temperature, where the thermal time scale diverges. A vibrational time scale which is well-defined in that limit is preferable. Noting that the definition of reduced units must satisfy the condition that the reduced quantity is still constant along isomorphs, we can define a new energy scale $e_1=e_0 \frac{h(\rho)}{k_BT} = h(\rho)$ and time scale $t_1 = t_0 \left(\frac{h(\rho)}{k_BT}\right)^{-1/2} = \rho^{-1/3}(h(\rho)/m)^{-1/2}$. These are independent of $T$ and therefore suitable for use in the limit $T\rightarrow0$. From the interpretation of $h(\rho)$ in terms of the curvature of the pair potential at the nearest neighbor distance\cite{Boehling/others:2014} we can interpret $t_1$ as a vibrational time scale for a single neighbor pair. Thus we can introduce an alternative reduced stress, denoted using a hat,

\begin{equation}
  \hat\sigma \equiv \frac{\sigma}{\rho h(\rho)}=\tilde\sigma \frac{k_BT}{h(\rho)}
\end{equation}
and alternative reduced strain rate
\begin{equation}
  \hat{\dot\gamma}\equiv \dot\gamma\rho^{-1/3}(h(\rho)/m)^{-1/2} = \tilde{\dot\gamma}\left(\frac{h(\rho)}{k_BT}\right)^{-1/2}.
\end{equation}
It is straightforward to re-write Eq.~\eqref{Chattoraj_flow_stress_v3} in terms of the alternative reduced units, giving

\begin{multline}\label{Chattoraj_flow_stress_v4}
  \hat\sigma(\dot\gamma, \rho, T) = \tilde A_0 + \tilde A_1\sqrt{\hat{\dot\gamma}} \\
  - \tilde A_2\left(\frac{k_BT}{h(\rho)}\right)^{2/3}[\ln(\tilde A_3\left(\frac{k_BT}{h(\rho)}\right)^{5/6}\hat{\dot\gamma}^{-1})]^{2/3}
\end{multline}
We thus recover an expression which resembles the original Eq.~\eqref{Chattoraj_flow_stress_v1} while still being explicitly isomorph invariant. We stress that the two expressions are equally valid, and that for the purpose of checking for isomorph invariance of a quantity the  choice of which system of reduced units is not important except for practical purposes (e.g. when $T=0$). However it can become relevant when comparing different isomorphs in order to identify the relevant physics, or for constructing a theory of the latter,  which is evident in the example above. Another example is the comparison of flow stress shown in Fig.~\ref{stress_stats}, where the strong temperature temperature of the reduced flow stress was partly ascribed to our choice of reduced units. Using $\rho h(\rho)$ instead of $\rho k_BT$ would probably reduce this variation, and is probably therefore more relevant for the glassy regime. Thus the advantages of one choice over the other are potentially greater clarity, insight, or ease of interpretation.

Lerner and Procaccia studied the flow stress for simulated glasses under steady state conditions covering both finite temperatures and the athermal limit~\cite{Lerner/Procaccia:2009a}, using a scaling theory based on the  approximation of their pair potential by an inverse power law. Noting that their exponent $\nu$ corresponds to our $\gamma+1$, all their scaling expressions are in fact compatible with isomorph theory, once one recognizes that their choice of time scaling is equivalent to our alternative reduced units. Their system is modeled using an approximate inverse power law potential which means that $h(\rho)$ is approximately a power law $\rho^\gamma$, in their notation $\rho^{\nu-1}$. Another example where the alternative choice of reduced stress was used was the athermal simulations of Ref.~\onlinecite{Lerner/Bailey/Dyre:2014} where the analysis was based on the isomorph theory and it was assumed (with little discussion) that the correct scaling of the stress at $T=0$ was $\rho h(\rho)$. Our point in the present discussion is that there is a choice of which system of reduced units to use, and that that choice is related to how relevant physics is best revealed. It is analogous to the choice of whether we study the standard diffusivity based on mean-squared displacement as a function of time, or the strain normalized diffusivity based on the mean squared displacement as a function of strain\cite{Chattoraj/Caroli/Lemaitre:2011}. We note again, however, that using $h(\rho)$\cite{Ingebrigtsen/others:2012, Boehling/others:2012}is less straightforward than $k_BT$ because it depends on the potential and is not directly available in the simulation. In some cases it is known analytically\cite{Ingebrigtsen/others:2012,Boehling/others:2014}, otherwise it must be identified from the shape of the numerically determined isomorph, before conversion into reduced units can take place.

\subsection{Improvements to future simulations}

Future work in this area could benefit from the following improvements.
(1) Our protocol assumes aging is negligible in our glassy undeformed systems, such that it makes sense to use generate isomorphs using fluctuations as if in equilibrium. It may be possible to avoid this assumption by using the fluctuations from the steady state shearing as the next best thing to equilibrium fluctuations. This possibility needs to be developed and evaluated theoretically.
(2) Another route to glassy isomorphs is to use the forces on particles in a single configuration, bypassing the need for equilibrium \cite{tbs_private:2019}.
(3) For comparing different isomorphs consistent reduced strain rates should be used, and moreover, different choices of how to define the reduced units should be considered, as discussed above.
Work along these lines is underway.


\section{\label{conclusion}Conclusion}

We have simulated isomorphs for the Kob-Andersen binary Lennard-Jones glass and compared their static structure and their dynamics under steady state shearing deformation. Two isomorphs were generated using the potential energy and virial fluctuations during and NVT simulation (no shear), assuming that aging effects could be ignored. This is probably a reasonable assumption for the lower temperature isomorph, but this is less clear for the high temperature one, which is only a few percent below the conventional mode-coupling temperature for this system, and therefore can be equilibrated as a liquid with longer (but still feasible) simulation times than we have used here. Nevertheless excellent collapse of the radial distribution function is observed, and good collapse for most of the dynamical measures. The worst collapse is observed for the shear stress autocorrelation function, which exhibited a systematic variation of the characteristic decay strain along an isomorph. Better statistics (i.e. longer runs) would probably help, but a more careful determination of the correct isomorph might be necessary, as it could be that this quantity is simply more sensitive to deviations from the correct isomorph than the others we have investigated. Going beyond simply checking for isomorph invariance we have analyzed the distributions of stress changes over different strain intervals. We showed that different features emerge according whether purely thermal effects are visible, or avalanches as indicated by an exponential tail in the distribution, or more complex and extremely non-Gaussian distributions at larger strain intervals which include multiple avalanches. Isomorph invariance is clear in all the data presented for this analysis. In comparing the mean squared transverse particle displacements, in addition to almost perfect isomorph invariance we noted how the MSD curves apparently become independent of temperature in the limit of long times, but also that one has to be careful to compare the same reduced strain rates. We note that no noticeable difference in the quality of the isomorphs is observed, despite the lower-temperature isomorph showing lower values of the correlation coefficient $R$ (see Table~\ref{tab:isomorph_data}). In the discussion, we showed how the existence of isomorphs can inform and constrain the development of analytical theories for how for example the flow stress can depend on density, temperature and strain rate. In addition there emerged an alternative definition of reduced units, the full implications of which will also be addressed in future work.

\begin{acknowledgments}
The work was supported in part by the VILLUM Foundation's {\em
Matter} Grant (No. 16515). This material is based upon work
supported by the National Science Foundation under Grant No.
CBET-1804186 (Y.J. and E.R.W.).
\end{acknowledgments}

\bibliography{Complete}

\end{document}